\documentclass[aps,prb,twocolumn,showpacs,floatfix]{revtex4-1}

\usepackage{verbatim}
\usepackage{graphicx}
\usepackage{dcolumn}
\usepackage{bm}
\usepackage{color}
\usepackage{hyperref}
\usepackage{makeidx}
\usepackage{amsmath}
\makeindex
\def\>{\right\rangle}
\def\<{\left\langle}
\def\be{\begin{equation}}
\def\ee{\end{equation}}
\def\ba{\begin{array}{l}}
\def\ea{\end{array}}

\def\beq{\begin{eqnarray}}
\def\eeq{\end{eqnarray}}

\begin{document}
\title{Time-resolved pure spin fractionalization and spin-charge separation in helical Luttinger liquid based devices}
\author{Alessio Calzona,$^1$ Matteo Carrega,$^2$ Giacomo Dolcetto,$^{2, 3}$ and Maura Sassetti$^{1,2}$}
\affiliation{ $^1$ Dipartimento di Fisica, Universit\`a di Genova,Via Dodecaneso 33, 16146, Genova, Italy.\\
$^2$ SPIN-CNR, Via Dodecaneso 33, 16146, Genova, Italy.\\
$^3$ Physics and Materials Science Research Unit, University of Luxembourg, L-1511 Luxembourg.
} 
\date{\today}
\begin{abstract}

Helical Luttinger liquids, appearing at the edge of two-dimensional topological insulators, represent a new paradigm of one-dimensional systems, where peculiar quantum phenomena can be investigated.
Motivated by recent experiments on charge fractionalization, we propose a setup based on helical Luttinger liquids that allows to time-resolve, in addition to charge fractionalization, also spin-charge separation and pure spin fractionalization. This is due to the combined presence of spin-momentum locking and interactions. We show that electric time-resolved measurements can reveal both charge and spin properties, avoiding the need of magnetic materials.
Although challenging, the proposed setup could be achieved with nowadays technologies, promoting helical liquids as interesting playgrounds to explore the effects of interactions in one dimension.
\end{abstract}
\maketitle
\section{Introduction}
Many body physics in low dimensions is a very active field in condensed matter~\cite{giamarchi2003quantum}. Here one-dimensional (1D) interacting systems play an important role~\cite{giamarchi2003quantum, deshpande2010electron, barak2010interacting}, since different fascinating phenomena have been predicted~\cite{giamarchi2003quantum, pham2000fractional, maslov1995landauer, safi1995transport} and observed~\cite{auslaender01042005, Lorenz2002evidence, bockrath1999luttinger, steinberg2007charge, kamata2014fractionalized}. The breakdown of the Fermi liquid paradigm in 1D led to the introduction of the Tomonaga-Luttinger liquid theory~\cite{tomonaga1950remarks, luttinger1963exactly}. This theory predicts peculiar phenomena, such as the spin-charge separation, {\it i.e.} interacting electrons split up into charge and spin collective excitations with different velocities~\cite{giamarchi2003quantum, deshpande2010electron}. This has been measured in semiconductor quantum wires~\cite{auslaender01042005}, carbon nanotubes~\cite{Lorenz2002evidence} and quasi-1D organic conductors~\cite{bockrath1999luttinger}.
Another intriguing aspect is charge fractionalization: an electron injected into an interacting region breaks up in collective excitations which carry fractional charges~\cite{pham2000fractional, maslov1995landauer, safi1995transport, tarucha1995reduction, safi1997properties, bena2001measuring, trauzettel2004appearance, berg2009fractional}. Although many theoretical works have concentrated on this, experimental evidences have been elusive for a long time. A first observation was achieved by Steinberg \textit{et al.} via momentum-resolved spectroscopy~\cite{steinberg2007charge, le2008charge}. More recently, Kamata {\it et al.} have reported the first direct measurement of fractional excitations by means of time-resolved transport measurements in a Hall bar setup~\cite{kamata2014fractionalized, perfetto2014time}.\\
In this work, we propose a setup in which, in addition to charge fractionalization and spin-charge separation, also pure spin fractionalization can be observed. This is not expected to occur in ordinary 1D liquids, due to spin-rotation invariance~\cite{safi1997properties}.
Our proposal is based on the edge states of two-dimensional (2D) topological insulators~\cite{roth2009nonlocal, hasan2010colloquium, qi2011topological}, whose properties are described in terms of counter-propagating channels characterized by spin-momentum locking. In the presence of interactions, these are described as helical Luttinger liquids (hLL)~\cite{wu06, dolcini11, dolcini12, romeo14, dolcetto2012tunneling, dolcetto13, das2011spin, Hou2009cornet, Schmidt2011current, Budich2012phonon, garate2012noninvasive}, whose first experimental manifestations have been recently observed~\cite{Li2015}. Here, charge and spin sectors are connected and the presence of interactions implies the simultaneous fractionalization of both degrees of freedom~\cite{das2011spin, calzona2015transient}. However, evidence of {\it pure} spin fractionalization, {\it i.e.} fractional spinons with no charge, have not yet been observed.
Indeed, the presence of measuring contacts prevents the observation of fractionalization via standard dc transport measurements~\cite{le2008charge, calzona2015transient}. Therefore, we propose time-resolved detection scheme, whose degree of control has been greatly improved~\cite{kamata2014fractionalized, ashoori1992edge, kumada2014intrinsic, waldie2015measurement}.
However, in hLL spin-momentum locking~\cite{wu06, dolcini11} prevents the independent observation of their fractionalization~\cite{giamarchi2003quantum}.
To circumvent this limitation, we consider a gate electrode to unbalance charge and spin velocities, allowing to separately identify fractionalized holons and spinons.
Our proposal relies {\it only} on a time-resolved electrical scheme, where also spin properties can be probed by means of conventional charge current measurements.
The proposed setup would allow to detect three fingerprints of interactions in the 1D world (spin-charge separation and their fractionalization) in a single experiment.
These results provide an important step forward in the understanding of the nature of interactions in one dimension.

\section{Model}
\begin{figure}[!ht]
\centering
\includegraphics[scale=1]{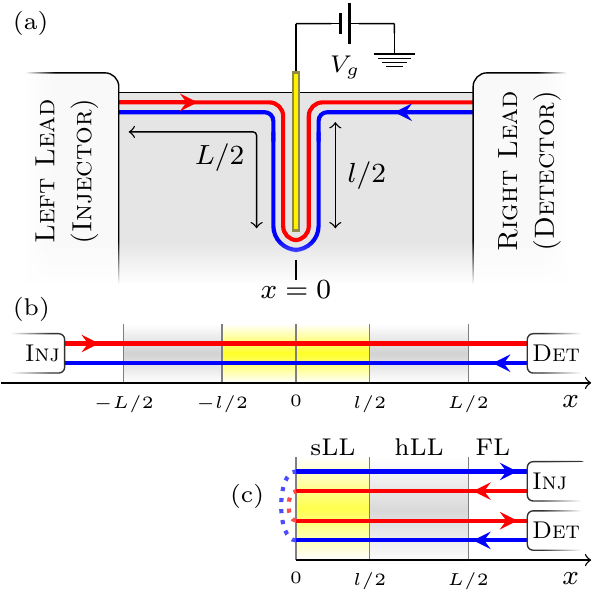}
\caption{(Color online) (a) Spin up (red) and spin down (blue) electrons of a single helical edge state counter propagate. Electron wave-packets can be pumped into the system from the injector and probed by the detector. An additional gate electrode (yellow) changes the edge state profile. (b) Model: the white regions correspond to the Fermi liquid leads, the gray ones to the edge states away from the gate electrode and the yellow ones to the edge states around the gate electrode. (c) Folding procedure applied to (b), that maps the infinite 2-channel (helical) liquid into the semi-infinite 4-channel (spinful) one.\\}\label{fig:1}
\end{figure}
We consider the setup in Fig. \ref{fig:1}(a)-(b), with helical edge modes of a 2D topological insulator described by
\begin{equation}\label{eq:H0}
\hat{\mathcal{H}}_0(x)=\hat{\psi}^{\dagger}_{\uparrow}(x)(-i\hbar v_F\partial_x)\hat{\psi}_{\uparrow}(x)+ \hat{\psi}^{\dagger}_{\downarrow}(x)(i\hbar v_F\partial_x)\hat{\psi}_{\downarrow}(x)
.\end{equation}
Short-range interactions are present, given by
\begin{equation}\label{eq:Heehelical}
\hat{\mathcal{H}}^{(L)} (x)=\frac{1}{2}\sum_{\sigma=\uparrow,\downarrow}\hat{\rho}_{\sigma}(x)\left [g_{4\parallel}(x)\hat{\rho}_{\sigma}(x)+g_{2\perp}(x)\hat{\rho}_{-\sigma}(x)\right ]
\end{equation}
with $\hat{\rho}_{\sigma}=\colon \hat{\psi}^{\dagger}_{\sigma}\hat{\psi}_{\sigma}\colon$ the electron particle density of each channel.
By negatively polarizing the central electrode the edge states are forced to change their profile.
Here we assume that the gate electrode is wide enough to prevent tunneling, but narrow enough to allow short-range interactions between the edge states on the two arms. This is indeed the case in the experiment performed by Kamata \textit{et al.} in Ref. \onlinecite{kamata2014fractionalized}.
Therefore the presence of the central electrode induces additional non-local interactions
\begin{equation}\label{eq:Heespinful}
\hat{\mathcal{H}}^{(l)} (x)=\frac{1}{2}\sum_{\sigma=\uparrow,\downarrow}\hat{\rho}_{\sigma}(-x)\left [g_{2\parallel}(x)\hat{\rho}_{\sigma}(x)+g_{4\perp}(x)\hat{\rho}_{-\sigma}(x)\right ]
.\end{equation}
The spatial dependence of the interaction parameters is $g_{4\parallel(2\perp)}(x)=g_{4\parallel(2\perp)}\Theta(L/2-|x|)$ and $g_{4\perp(2\parallel)}(x)=g_{4\perp(2\parallel)}\Theta(l/2-|x|)$.
The edges are connected to Fermi-liquid metallic contacts, modeled as 1D non-interacting systems~\cite{maslov1995landauer, safi1995transport}.
Note that a rigorous microscopic calculation of the dependence of the strength of interactions on the geometric gate parameters would involve the full numerical solution of a bidimensional Poisson problem posed by the whole setup, which is beyond the scope of the present article.
Nevertheless, modelling the effect of the gate voltage on the edge state physics through short-range Coulomb interactions in the form of Eq. \ref{eq:Heespinful} has been shown to be a very good effective description, corroborated by experimental results~\cite{kamata2014fractionalized}.\\
Interactions are treated within bosonization~\cite{giamarchi2003quantum}. Introducing the scalar fields $\hat{\phi}_{\sigma}$ related to the electron density as $\hat{\rho}_{\uparrow/\downarrow}(x)=\mp\frac{1}{\sqrt{2\pi}}\partial_x\hat{\phi}_{\uparrow/\downarrow}(x)$
the Hamiltonian is diagonalized in terms of the fields
\begin{equation}\label{eq:trasf}
\left( \begin{array}{cc} \phi_c(x) \\ \theta_c(x) \\ \phi_s(x) \\ \theta_s(x) \end{array} \right)=\frac{1}{2}\left( \begin{array}{cccc} -1 & 1 & 1 & -1 \\ 1 & 1 & 1 & 1 \\ -1 & 1 & -1 & 1 \\ 1 & 1 & -1 & -1 \end{array} \right)\left( \begin{array}{cc} \phi_{\uparrow}(x) \\ \phi_{\uparrow}(-x) \\ \phi_{\downarrow}(x) \\ \phi_{\downarrow}(-x) \end{array} \right)
,\end{equation}
satisfying $[\partial_x\hat{\phi}_{\lambda}(x),\hat{\theta}_{\lambda '}(x')]=i\delta_{\lambda,\lambda '}\delta(x-x')$. The bosonized version of $\hat{H}=\int_{-\infty}^{\infty}~dx~[\hat{\mathcal{H}}_0(x)+\hat{\mathcal{H}}^{(L)} (x)+\hat{\mathcal{H}}^{(l)} (x)]$ can be written as $\hat{H}=\sum_{\lambda=c,s}\hat{H}_{\lambda}$, with
\begin{equation}\label{eq:Hll}
\hat{H}_{\lambda}=\int_0^{\infty}dx\frac{v^{(\lambda)}(x)}{2}\left \{\frac{\left [\partial_x\hat{\phi}_{\lambda}(x)\right ]^2}{K^{(\lambda)}(x)}+K^{(\lambda)}(x)\left [\partial_x\hat{\theta}_{\lambda}(x)\right ]^2\right \}
.\end{equation}
The transformations Eq. (\ref{eq:trasf}) fictitiously double the degrees of freedom of the system, whose correct number is restored by halving the domain of integration. The infinite 2-channel liquid with non-local interactions, Fig. \ref{fig:1}(b), is mapped onto the semi-infinite 4-channel liquid with local interactions~\cite{g-ology}, Fig. \ref{fig:1}(c).
Equations (\ref{eq:trasf}) define charge and spin fields of the spinful-like 4-channel liquid, related to its charge (units $e$) and spin (units $\hbar/2$) densities and currents as $\hat{\rho}_{\lambda}=\sqrt{2/\pi}\partial_x\hat{\phi}_{\lambda}$ and $\hat{j}_{\lambda}=-\sqrt{2/\pi}\partial_t\hat{\phi}_{\lambda}$.
Equation (\ref{eq:Hll}) represents the Hamiltonian for a spinful Luttinger liquid (sLL). Charge and spin sector separates, characterized by velocities
\begin{equation}\label{eq:velocities}
v^{(\lambda)}(x)=v_F\sqrt{\left (1+\bar{g}_{4,\lambda}(x)\right )^2-\bar{g}_{2,\lambda}^2(x)}
\end{equation}
and Luttinger parameters
\begin{equation}\label{eq:LuttingerK}
K^{(\lambda)}(x)=\sqrt{\frac{1+\bar{g}_{4,\lambda}(x)-\bar{g}_{2,\lambda}(x)}{1+\bar{g}_{4,\lambda}(x)+\bar{g}_{2,\lambda}(x)}}
,\end{equation}
with $\bar{g}_{i,c}(x)=\frac{g_{i\parallel}(x)+g_{i,\perp}(x)}{2\pi v_F}$ and $\bar{g}_{i,s}(x)=\frac{g_{i,\parallel}(x)-g_{i,\perp}(x)}{2\pi v_F}$.
We refer to $l/2<x<L/2$ as the hLL-region, and  sLL-region the one for $0\leq x<l/2$, see Fig. \ref{fig:1}(c).
As shown in Table \ref{table:parameters}, inhomogeneous interactions generate space-dependent parameters.
\begin{table}
\begin{center}
    \begin{tabular}{ | c | c | c | c | p{3cm} |}
    \hline
     & sLL-region & hLL-region & FL-region \\
& $0<x<l/2$ & $l/2<x<L/2$ & $L/2<x$ \\
\hline
    $K^{(c)}(x)$ & $K_c$ & $K_h$ & $1$ \\ \hline
    $K^{(s)}(x)$ & $K_s$ & $1/K_h$ & $1$ \\ \hline
    $v^{(c)}(x)$ & $v_c$ & $v_h$ & $v_F$ \\ \hline
    $v^{(s)}(x)$ & $v_s$ & $v_h$ & $v_F$ \\ 
    \hline
    \end{tabular}
\end{center}
\caption{Luttinger parameters and velocities appearing in the inhomogeneous Luttinger liquid model Eqs. (\ref{eq:Hll})-(\ref{eq:LuttingerK}), corresponding to the setup of Fig. \ref{fig:1}.}
\label{table:parameters}
\end{table}
This inhomogeneous Luttinger liquid model has been applied to explain conductance quantization in interacting wires ~\cite{maslov1995landauer, safi1995transport, safi1997properties} and charge fractionalization in quantum Hall devices~\cite{kamata2014fractionalized, perfetto2014time, Agarwal2014time, ferraroprl14}.
The setup of Fig. \ref{fig:1}(a) represents a particular inhomogeneous Luttinger liquid, where Fermi liquid (FL), hLL and sLL behaviors coexist. At the FL-hLL interface (hLL$|$FL) at $x=L/2$ and at the hLL-sLL interface (sLL$|$hLL) at $x=l/2$, the inhomogeneous interactions affect the propagation of excitations.

\section{Time evolution}
We  study space-time evolution of electron wave-packets injected from the left contact (injector) and detected by the right one (detector).
Injection can be achieved by {\it e.g.} applying a voltage pulse to the injection gate, while quantum point contacts can be exploited to perform time-resolved detection~\cite{kamata2014fractionalized, waldie2015measurement, bocquillon2012electron, bocquillon2014electron, wahl13, ferraro14}.
Crucially, the concept of \textit{electron} wave-packet makes sense in the FL-region only: once entered the interacting region, the electron wave-packet is decomposed into charge and spin collective excitations, governed by the equations of motion~\cite{safi1995transport, maslov1995landauer, dolcini2005transport}
\begin{equation}\label{eq:4motion}
\partial^2_t\hat{\phi}_{\lambda}(x,t)=v^{(\lambda)}(x)K^{(\lambda)}(x)\partial_x\left [\frac{v^{(\lambda)}(x)}{K^{(\lambda)}(x)}\partial_x\hat{\phi}_{\lambda}(x,t)\right ]
,\end{equation}
obtained by recalling $\partial_t\hat{\phi}_{\lambda}(x,t)=v^{(\lambda)}(x)K^{(\lambda)}(x)\partial_x\hat{\theta}_{\lambda}(x,t)$ and $\partial_t\hat{\theta}_{\lambda}(x,t)=\frac{v^{(\lambda)}(x)}{K^{(\lambda)}(x)}\partial_x\hat{\phi}_{\lambda}(x,t)$ .
From Eq. (\ref{eq:4motion}) one can determine the dynamical evolution of $\phi_{\lambda}(x,t)\equiv\langle\hat{\phi}_{\lambda}(x,t)\rangle$, by imposing the continuity of $\phi_{\lambda}(x,t)$ and $\frac{v^{(\lambda)}(x)}{K^{(\lambda)}(x)}\partial_x\phi_{\lambda}(x,t)$ in $x=l/2,L/2$, together with $\partial_t \phi_{\lambda}(0,t)=0$, the latter imposing total reflection~\cite{note} at $x=0$ (see Fig. \ref{fig:1}(c)).
The equation of motion can be mapped into a scattering problem for an incident plasmon mode, as described in Appendix~\ref{appendix}.
Space and time derivatives of $\phi_{\lambda}(x,t)$ are related to the average densities $\rho_{\lambda}(x,t)$ and currents $j_{\lambda}(x,t)$ respectively.
To determine their dynamical evolution, the initial ($t=0$) density $\rho^{(0)}_{\lambda}(x)$ and current $j^{(0)}_{\lambda}(x)$ profiles, determined by the injection process, must be specified.
Because of helicity, only spin-up electrons can propagate from the injector. Therefore we consider $\rho^{(0)}_{c}(x)=\rho^{(0)}_s(x)\equiv\rho^{(0)}(x)$ and $j^{(0)}_c(x)=j^{(0)}_s(x)=v_F\rho^{(0)}(x)$, with $\rho^{(0)}(x)=N\exp[(x-x_{i})^2/2\sigma^2]/(\sqrt{2\pi}\sigma)$ assumed to have a Gaussian distribution, $x_i>L/2$ being the injection point.
$N$ is the number of injected electrons, with total charge $Q$ and spin $S=\hbar N/2$. The parameter $\sigma$ can describe the finite duration of the injection process; it is related to the full-width-at-half-maximum $\delta t$ of the time distribution of the injected wave-packets as $\sigma\approx 0.42v_F\delta t$.
By solving Eq. (\ref{eq:4motion}), the dynamical evolution of charge and spin sectors can be determined (see Appendix \ref{appendix}).
This can be written as $\rho_{\lambda}(x,t)=\rho^{(inc)}_{\lambda}(x,t)+\rho^{(refl)}_{\lambda}(x,t)$, with $\rho^{(inc)}_{\lambda}(x,t)$ the wave-packet incoming from the injector and $\rho^{(refl)}_{\lambda}(x,t)$ the wave-packet reflected to the FL-region (see Fig. \ref{fig:1}(c)).
The latter encodes information about scattering phenomena and fractionalizations, and is related to the detected current (units $e$) as
\begin{equation}
j_{det}(x_d,t)=\frac{v_F}{2}\left [\rho^{(refl)}_c(x_d,t)+\rho^{(refl)}_s(x_d,t)\right ]\label{eq:j}
,\end{equation}
with $x_d>L/2$ the detection point.
Crucially, due to helicity, information about both charge and, more importantly, spin can be extracted from the detected current.
By recalling Eqs. (\ref{eq:suppAc}-\ref{eq:supprho(x,t)}) one has
\begin{widetext}
\begin{equation}
\rho^{(refl)}_c(x_d,t)=r\rho^{(0)}(-x_d+v_Ft+L)+\gamma\sum_{n,m=0}^{\infty}C_{n,m}\rho^{(0)}\left (-x_d+v_Ft+L-(n+1)\frac{v_F}{v_h}(L-l)-m\frac{v_F}{v_c}l\right ),\label{eq:rho(x,t)}
\end{equation}
\begin{equation}
C_{n,m}=\left (\frac{1-K_h}{1+K_h}\right )^n\left (\frac{K_c-K_h}{K_c+K_h}\right )^{n+m+1}\left [(1-\delta_{m,0})\sum_{p=1}^{\mathrm{min}\{m,n+1\}}{{m-1}\choose{p-1}}{{n+1}\choose{p}}\left [\frac{-4 K_hK_c}{\left (K_c-K_h\right )^2}\right ]^p+\delta_{m,0}\right ]\label{eq:Ac}
,\end{equation}
\end{widetext}
with $r=\frac{1-K_h}{1+K_h}$, $\gamma=-\frac{4K_h}{(1+K_h)^2}$. The spin density evolution $\rho^{(refl)}_s(x_d,t)$ can be obtained from Eqs. (\ref{eq:rho(x,t)}-\ref{eq:Ac}) with the substitutions
\begin{equation}\label{eq:substitution}
\left (v_c,K_c,K_h\right )\to\left (v_s,K_s,1/K_h\right ).
\end{equation}
Note that in the non-interacting case $\rho^{(refl)}_c(x_d,t)=\rho^{(refl)}_s(x_d,t)=\rho^{(0)}(-x_d+v_Ft)$: the injected wave-packet simply propagate through the system with constant velocity $v_F$, and does not undergo neither fractionalizations nor spin-charge separation (see inset of Fig. \ref{fig:2}).\\
The first term in the r.h.s. of Eq. (\ref{eq:rho(x,t)}) corresponds to the injected wave-packet being reflected at hLL$|$FL. Because of helicity, the reflection of a fraction $rQ$ of the injected charge $Q$ is accompanied by a reflected fraction $-rS$ of the total incoming spin $S$.
Spin fractionalization is due to the helical nature of the edge states, which, in the presence of interactions, effectively breaks the spin-rotation symmetry.
The component transmitted across hLL$|$FL generates the second term of Eq. (\ref{eq:rho(x,t)}), where an infinite number of terms appears.
Once entered the interacting region, the transmitted wave-packet can undergo multiple reflections between the interfaces before going back to the FL-region.
Terms proportional to $C_{n,m}$ correspond to $n$ reflections at hLL$|$FL toward hLL-region and $m$ reflections at $x=0$~(see Appendix~\ref{appendix}).\\
Note that for long times $\int_0^{\infty} dt j^{(refl)}_c(x_d,t)=Q$ and $\int_0^{\infty} dt j^{(refl)}_s(x_d,t)=S$, with $j^{(refl)}_{\lambda}(x_d,t)=v_F\rho^{(refl)}_{\lambda}(x_d,t)$. Then, from Eq. (\ref{eq:j}), $\int_0^{\infty} dt j_{det}(x_d,t)=Q$: in the dc limit, the injected wave-packet is transmitted to the detector, in agreement with the universal quantization of dc conductance~\cite{safi1995transport, safi1997properties}.

\section{Results and discussion}
\begin{figure}[!ht]
\centering
\includegraphics[scale=0.38]{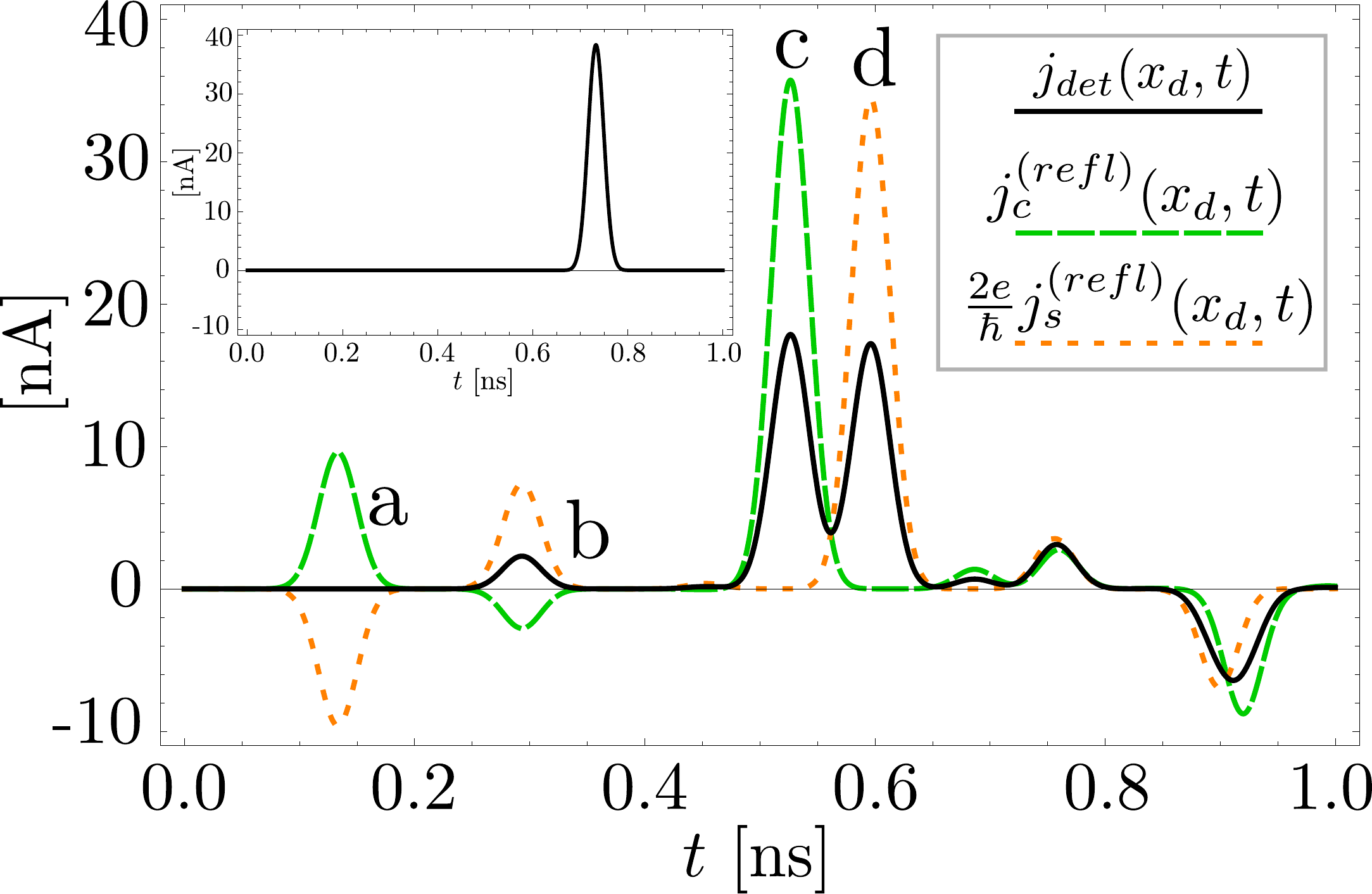}
\caption{(Color online) Detected, charge and spin (multiplied by $2e/\hbar$) currents as a function of time.
The injected wave-packet consists of $N=10$ spin-up-polarized electrons, whose time distribution has a width-at-half-maximum $\delta t\approx 40$ ps.
Parameters are $x_i=x_d=12$ $\mu$m, $L=18$ $\mu$m, $l=10$ $\mu$m. The Luttinger parameters~\cite{note2} are $v_c=v_F/K_c$, $v_h=v_F/K_h$, $v_s=v_FK_s$, $v_F=3\times 10^4$ m/s and $K_h=0.6$, $K_c=0.7$, $K_s=1.1$. The inset shows the detected current without interactions, where neither fractionalizations nor spin-charge separation are present.}\label{fig:2}
\end{figure}

\begin{figure}[!ht]
\centering
\includegraphics[scale=1]{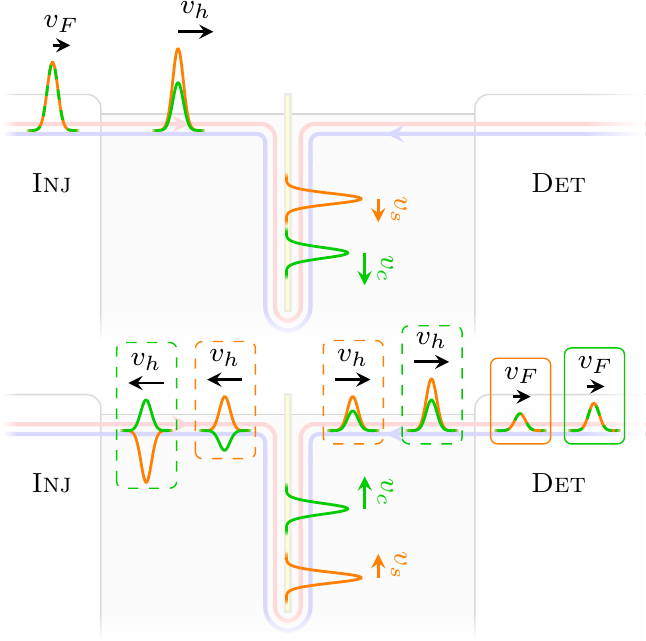}
\caption{(Color online) Physical processes corresponding to signals c and d in Fig. \ref{fig:2}. Charge and spin excitations are represented in green and orange respectively. (top) The charge and spin components of the injected wave-packet are fractionalized when entering the interacting region. Here they propagate with the same velocity $v_h$ reaching the gate electrode, where they are transmitted and further fractionalized. Around the electrode charge and spin separate ($v_c>v_s$). (bottom) The charge (spin) component returns back and is split into two wave-packets, represented by dashed green (orange) rectangles, with opposite spin (charge), one propagating toward the detector and the other one toward the injector. The components propagating toward the detector are finally probed, with a time shift $\Delta t$, as electric signals in the time-resolved current (solid rectangles).}\label{fig:3}
\end{figure}

The detected current as a function of time $j_{det}(x_d,t)$ is shown, with realistic parameters, in Fig. \ref{fig:2}, together with charge and spin currents of the 4-channel liquid.
Different signals are revealed at different times, reflecting signatures of fractionalization.
We focus on the signals a-d, that are captured by the short-time contributions of Eq. (\ref{eq:rho(x,t)}).
For the charge sector they read
\begin{equation}\label{eq:rho}
\begin{split}
\rho^{(refl)}_c&(x,t)\approx r\rho^{(0)}(-x+v_Ft+L)\\
&+\gamma C_{0,0}\rho^{(0)}\left (-x+v_Ft+L-v_F\frac{L-l}{v_h}\right )\\
&+\gamma C_{0,1}\rho^{(0)}\left (-x+v_Ft+L-v_F\frac{L-l}{v_h}-v_F\frac{l}{v_c}\right )
,\end{split}
\end{equation}
and similarly for the spin sector, by using Eq. (\ref{eq:substitution}).
The first charge and spin components, first member in the r.h.s. of Eq. \eqref{eq:rho}, corresponds to the injected wave-packet being reflected before entering the interacting region, and thus are not probed by the detector (signals a).
The components transmitted across hLL$|$FL propagate in the hLL-region as charge and spin density waves.
At sLL$|$hLL they are partially reflected back, second member in the r.h.s. of Eq. (\ref{eq:rho}), with charge $\frac{4K_h(K_h-K_c)}{(1+K_h)^2(K_c+K_h)}Q$ and spin $\frac{4K_h(1-K_hK_s)}{(1+K_h)^2(K_hK_s+1)}S$, leading to a finite detected current (peak b).
However, this signal does not allow to independently identify charge and spin fractionalization: the current Eq.~(\ref{eq:j}) is a superposition of both fractional charge and spin.\\
The transmitted components across sLL$|$hLL, third member in the r.h.s. of Eq. (\ref{eq:rho}), produce signals c and d in Fig. \ref{fig:2}.
The corresponding physical processes are shown in Fig. \ref{fig:3}.
Once the wave-packet enters the sLL-region, its charge and spin components separate, with charge density waves propagating faster than spin ones ($v_c>v_s$), as shown in Fig. \ref{fig:3}(top). Therefore the charge excitation returns back to sLL$|$hLL before the spin one.
Here the charge (spin) wave-packet is split into two wave-packets with opposite spin (charge), marked by dashed green (orange) rectangles in Fig. \ref{fig:3}(bottom).
The components propagating toward the detector are finally probed as electric signals c and d in Fig. \ref{fig:2}, with a time delay $\Delta t=\frac{v_c-v_s}{v_cv_s}l$ due to the different time of flights of pure charge and pure spin excitations.
This time shift represents a manifestation of spin-charge separation. The peak marked with c in Fig. \ref{fig:2} is due to fractionalization of pure charge excitations (holons), while the peak marked with d is related to fractionalization of pure spin (spinons).
The other signals observed in Fig. \ref{fig:2} are due to several reflections inside the interacting region.
Note that, for different interaction parameters, the precise structure of the current can change, but the signals reflecting pure charge and pure spin fractionalization can be identified in any case.

Therefore, we come to the main result of our work. The proposed setup allows to observe the landmarks of interactions in one dimension by means of a unique electric time-resolved detection scheme: fractionalization of holons and spinons can be identified, the time delay in their detection providing manifestation of spin-charge separation.
This is particularly fascinating for the possibility to electrically probe pure spin fractionalization.

\section{Experimental feasibility}
Key point of the proposal is the ability to time-resolve the charge and spin fractionalization, signals c and d in Fig. \ref{fig:2}.
This implies $\Delta t>\delta t$, that is, the spin-charge time shift $\Delta t$ greater than the width of the time distribution of the wave-packets $\delta t$.
The ability to inject electron wave-packets with $\delta t$ of the order of tens of ps~\cite{waldie2015measurement}, together with realistic values of the Fermi velocity ($v_F\approx 3\times 10^4$ m/s for the edge states in InAs/GaSb quantum wells), gives a lower bound on the system size of approximatively $1$ $\mu$m.
An upper bound is given by the inelastic mean free path $l_{\mathrm{in}}$. Ballistic transport has been observed up to few $\mu$m-long samples, but further comprehension of the scattering mechanisms~\cite{vayrynen2013helical, vayrynen2014resistance} would allow to increase the coherence length up to tens of $\mu$m~\cite{garate2012noninvasive,konig2007quantum, konig2008quantum, konig2013spatially, gusev2014temperature, olshanetsky2015persistence, essert2015two}.
The plot of Fig. \ref{fig:2} corresponds to $\sim 10$ $\mu$m-long setups, and the charge and spin signals, although slightly overlapping, can be resolved.
Therefore, the proposed measurements can be likely performed with current technologies, and helical edge states reasonably represent an promising playground to explore the effects of interactions in one dimension.
\begin{acknowledgments}
We acknowledge the support of the MIUR-FIRB2012 - Project HybridNanoDev (Grant  No.RBFR1236VV), EU FP7/2007-2013 under REA grant agreement no 630925 -- COHEAT, MIUR-FIRB2013 -- Project Coca (Grant No.~RBFR1379UX), the COST Action MP1209 and the CNR-SPIN via Seed Project PGESE003. G.D. is supported by the National Research Fund, Luxembourg under grant ATTRACT 7556175.
\end{acknowledgments}


\appendix

\section{Time evolution of charge and spin density}\label{appendix}
Here we concentrate on the time evolution of the average quantities $\phi_\lambda (x,t)=\langle \hat{\phi}_\lambda (x,t)\rangle$ which satisfy the equations of motion
\begin{equation}
\label{eq:motion}
\partial_t^2 \phi_\lambda(x,t) = v^{(\lambda)}(x) K^{(\lambda)}(x) \partial_x \left[\frac{v^{(\lambda)}(x)}{K^{(\lambda)}(x)} \partial_x \phi_\lambda(x,t)\right].
\end{equation}
The parameters $v^{(\lambda)}(x)$ and $K^{(\lambda)}(x)$ are piecewise constants, being discontinuous at the junctions between different regions.
To evaluate the densities $\rho_{\lambda}(x,t)=\sqrt{2/\pi}\partial_x\phi_{\lambda}(x,t)$ and currents $j_{\lambda}(x,t)=-\sqrt{2/\pi}\partial_t\phi_{\lambda}(x,t)$ it is necessary to evaluate the space-time evolution of $\phi_{\lambda}(x,t)$. This can be done solving Eq. (\ref{eq:motion}) by imposing that $\phi_{\lambda}(x,t)$ and $\frac{v^{(\lambda)}(x)}{K^{(\lambda)}(x)}\partial_x \phi_{\lambda} (x,t)$ are continuous functions and that the current $j_{\lambda}(0,t)=0$.
Due to the linearity in Eq. (\ref{eq:motion}) $\phi_{\lambda}(x,t)$ can be decomposed into bosonic modes
\begin{equation}
\label{eq:phi_s}
\phi_{\lambda,q}  (x,t) = e^{-iv_F q t} s_{\lambda,q}(x)
\end{equation}
with momentum $q$ and energy $\hbar v_Fq$ as
\begin{equation}\label{eq:decomposition}
\phi_{\lambda}(x,t) = \int_{-\infty}^{+\infty} dq \, a_{\lambda,q} \, \phi_{\lambda,q}(x,t)
,\end{equation}
the coefficients $a_{\lambda,q}$ being related to the distribution of the injected  wave-packet, as will be shown in the following.
Therefore, the problem of solving the equation of motion for $\phi_{\lambda}(x,t)$ is mapped into a scattering problem for the incident mode $\phi_{\lambda,q}(x,t)$ in Eq. (\ref{eq:phi_s}), in agreement with the generalized plasmon scattering approach.\\
In the following we focus on the charge sector only, and omit the index $\lambda=c$ for notational convenience (for example $s_q(x)\equiv s_{\lambda , q}(x)$); the spin sector with $\lambda=s$ can be obtained with the substitutions $(v_c,K_c,K_h) \rightarrow (v_s,K_s,1/K_h)$, as reported in Table I of the main text.
For the charge sector
\begin{eqnarray}
v(x) &=& \begin{cases}
v_F & x>L/2 \\
v_h & l/2<x<L/2 \\
v_c & 0\leq x< l/2
\end{cases}\\
K(x) &=& \begin{cases}
1 & x>L/2 \\
K_{h} & l/2<x<L/2 \\
K_{c} & 0\leq x< l/2
\end{cases}
.\end{eqnarray}
Therefore to solve the scattering problem one requires that
\begin{equation}
\label{eq:s}
s_{q} (x) = \begin{cases}
e^{-i q x} + R_{q} e^{iq x} & x>L/2\\
A_{q} e^{-iq\frac{v_F}{v_h}x} + B_{q} e^{iq\frac{v_F}{v_h}x} & l/2<x<L/2\\
C_{q} e^{-iq\frac{v_F}{v_c}x} + D_{q} e^{iq\frac{v_F}{v_c}x} & 0\leq x<l/2
\end{cases}~,
\end{equation}
and $\frac{v(x)}{K(x)}\partial_xs_{q}(x)$ are continuous functions at $x=l/2,L/2$, together with $s_q(0)=0$ which guarantees $j(0,t)=0$.
With these constrains it is possible to determine $R_q$, $A_q$, $B_q$, $C_q$ and $D_q$. In particular, we are interested to evaluate $R_q$, which is related to the components of the injected  wave-packet entering the detector (in the four-channel model of Fig. 1(c) in the main text, only reflected components of the injected  wave-packet can be probed by the detector).
Once $R_q$ is known, we can use Eq. (\ref{eq:decomposition}) to get, for $x>L/2$,
\begin{eqnarray}
\label{eq:sviluppo}
\rho(x,t)&=&\sqrt{\frac{2}{\pi}}\int_{-\infty}^{+\infty} dq~a_q \, e^{-iqv_F t} \partial_x \left(e^{-iqx} + R_q e^{iqx} \right)\nonumber\\
&\equiv &\rho^{(inc)}(x,t)+\rho^{(refl)}(x,t)
,\end{eqnarray}
where the incoming
\begin{equation}
\label{eq:sviluppo_inc}
\rho^{(inc)}(x,t)=-i\sqrt{\frac{2}{\pi}}\int_{-\infty}^{+\infty} dq~a_q q \, e^{-iq(x+v_F t)}
\end{equation}
and reflected
\begin{equation}
\label{eq:sviluppo_refl}
\rho^{(refl)}(x,t)=i\sqrt{\frac{2}{\pi}}\int_{-\infty}^{+\infty} dq~a_q qR_q \, e^{iq(x-v_F t)}
\end{equation}
components have been identified.
At $t=0$ the incoming  wave-packet is specified by the injection process, \textit{i.e.} $\rho^{(inc)}(x,0)=\rho^{(0)}(x)$, which is non-vanishing close to the injection point $x_i$ in the FL region.
By comparing Eq. (\ref{eq:sviluppo_inc}) with the Fourier transform of the injected wave-packet
\begin{equation}
\label{eq:sviluppo_inj}
\rho^{(0)}(x)=\int_{-\infty}^{+\infty} \frac{dq}{2\pi}~\rho^{(0)}_q \, e^{-iqx}
,\end{equation}
one finds $a_q=\frac{i}{\sqrt{2\pi}2q}\rho^{(0)}_q$ and from Eq. (\ref{eq:sviluppo_refl})
\begin{equation}
\label{eq:riflesso}
\rho^{(refl)}(x,t)=-\int_{-\infty}^{+\infty} \frac{dq}{2\pi}~\rho^{(0)}_qR_q \, e^{iq(x-v_F t)}
.\end{equation}
By means of Eq. (\ref{eq:riflesso}) it is possible to determine the evolution of the reflected component from the knowledge of $\rho^{(0)}_q$, related to the shape of the injected  wave-packet, and from $R_q$, related to the solution of the scattering problem, that is explicitly determined in the following.

\subsection{Reflection coefficient $R_q$}
\begin{figure}
\centering
\includegraphics[width=1\linewidth]{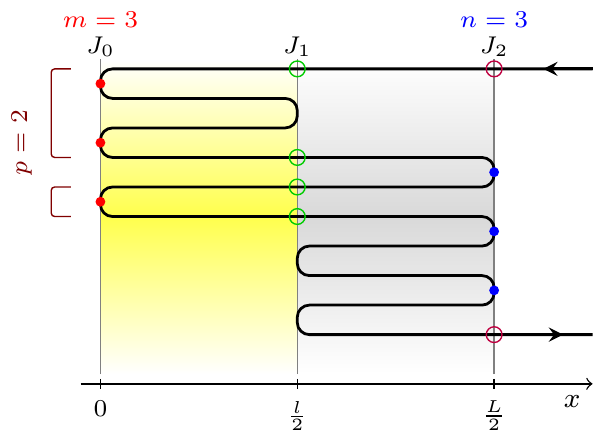}
\caption{(Color online) An example of a possible path contributing to the reflection coefficient $R_q$. After the injected wave packet is transmitted across $J_2$ to the interacting region (upper purple circle) and before it is reflected back to the FL-region $x>L/2$ (lower purple circle), it undergoes multiple reflections.
In particular, the specific path illustrated here is characterized by $n=3$ reflections at $J_2$ (blue dots) and $m=3$ reflections at $J_0$ (red dots). There are $p=2$ disconnected subpaths inside the sLL-region $x<l/2$. Therefore there are $2p=4$ transmissions across $J_1$ (green circles), $m-p=1$ reflection from $J_1$ toward $J_0$ and $n+1-p=2$ reflections from $J_1$ toward $J_2$, in agreement with Eq. (\ref{eq:Pnm}).}
\label{fig:Supple}
\end{figure}

Solving the scattering problem is, in principle, straightforward: the five coefficients appearing Eq. (\ref{eq:s}) can indeed be determined from the five boundaries conditions. However this approach is not convenient from a physical point of view. Indeed, it does not allow to discriminate the different physical processes occurring in the system, due to multiple reflections of the injected  wave-packet.
To overcome this limitation and be able to follow the space-time evolution of the collective excitations, we adopt a different strategy.
We consider all possible paths that a bosonic mode Eq. (\ref{eq:phi_s}), once entered the interacting region $x<L/2$, can follow before being reflected back in the region $x>L/2$. A possible example of one of these paths is shown in Fig. \ref{fig:Supple}.
We call $J_2$ the hLL$|$FL junction at $x=L/2$, $J_1$ the sLL$|$hLL junction at $x=l/2$ and $J_0$ the end of the whole (folded) system at $x=0$.\\
The first contribution to the reflection coefficient $R_q$ is obviously due to the first reflection from $J_2$: the injected wave-packet does not enter the interacting region. We denote this contribution as $r_{q,\leftarrow}^{(J_2)}$, the arrow $\leftarrow$ ($\rightarrow$) denoting incoming wave-packet incident from the right (left). As for the other coefficients that will be introduced in the following, we refer to the next section for its explicit evaluation.
The other contributions are due to the fraction of the injected wave-packet transmitted across $J_2$ into the interacting region $x<L/2$. Here the fractionalized wave packet can undergo several reflections from $J_0$, $J_1$ and $J_2$ before going back to $x>L/2$, where it can be probed by the detector.
All these contributions pick up a factor $t_{q,\rightarrow}^{(J_2)} t_{q,\leftarrow}^{(J_2)}$, \textit{i.e.} the product of the transmission coefficient across $J_2$ when entering ($t_{q,\leftarrow}^{(J_2)}$) and exiting ($t_{q,\rightarrow}^{(J_2)}$) the interacting region (see upper and lower purple circles in Fig. \ref{fig:Supple} respectively).
Note that, once entered the interacting region $x<L/2$, the wave packet arrives at $J_1$ and, just before exiting the interacting region, it comes from $J_1$ again (see Fig. \ref{fig:Supple}). In order to consider all the contributions to $R_q$ we conveniently focus on the paths that start and arrive at $J_1$.\\
Let $n$ be the number of reflections at $J_2$ toward $J_1$ and  $m$ the number of reflections at $J_0$ (toward $J_1$ obviously). Both $n$ and $m$ can assume natural values. For the specific case shown in Fig. \ref{fig:Supple} $n=3$ and $m=3$.
The reflection coefficient from $J_2$ toward $J_1$ is $r_{q,\rightarrow}^{(J_2)}$, while with $r_{q,\leftarrow}^{(J_0)}$ we denote the reflection coefficient from $J_0$.
The contributions to $R_q$ due to several reflections is thus given by $\left (r_{q,\rightarrow}^{(J_2)}\right )^n$ and $\left (r_{q,\leftarrow}^{(J_0)}\right )^m$.
Therefore the global reflection coefficient can be written as
\begin{equation}
\label{eq:R}
R_q = r_{q,\leftarrow}^{(J_2)} + t_{q,_\rightarrow}^{(J_2)} t_{q,\leftarrow}^{(J_2)} \sum_{n,m=0}^{+\infty} \left(r_{q,\leftarrow}^{(J_0)}\right)^m  \left(r_{q,\rightarrow}^{(J_2)}\right)^n P_{n,m}
.\end{equation}
The coefficient $P_{n,m}$ takes into account the contribution of all the different paths with fixed $m$ and $n$. Indeed, assigning $(n,m)$ does not completely specify the path.
Moreover, also transmission and reflections at $J_1$ contribute to the coefficient of a specific path, and are encoded into $P_{n,m}$.
Let us see this point in more detail.\\
If $m=0$ there are no reflections from $J_0$ and $P_{n,0}$ is given by $\left (r_{q,\leftarrow}^{(J_1)}\right )^{n+1}$, \textit{i.e.} the $n+1$ reflections at $J_1$ toward $J_2$. The calculation is more complicated if $m>0$.
It is useful to introduce the integer number $p$, that counts the number of disconnected subpaths inside the sLL-region ($x<l/2$). For example, $p=2$ in Fig. \ref{fig:Supple}.
The number $p$ is bounded as $1\leq p \leq \min\{m,n+1\}$.
If there are $p$ disconnected subpaths in the sLL-region, then there are $p$ transmission across $J_1$ from the hLL-region to the sLL-region $\left(t_{q,\leftarrow}^{(J_1)}\right)^p$, $p$ transmissions across $J_1$ from the sLL-region to the hLL-region $\left(t_{q,\rightarrow}^{(J_1)}\right)^p$, $(m-p)$ reflections at $J_1$ toward $J_0$ $\left({r_{q,\rightarrow}^{(J_1)}}\right)^{m-p}$ and $(n+m+1)-2p-(m-p) = n+1-p$ reflections at $J_1$ toward $J_2$ $\left({r_{q,\leftarrow}^{(J_1)}}\right)^{n+1-p}$.\\
Finally, the grouping of the $m$ and $n$ reflections into $p$ subpaths is not unique, so that additional combinatory factors must be kept into account.
The final expression for $P_{n,m}$ reads
\begin{widetext}
\begin{equation}
\label{eq:Pnm}
P_{n,m} = \delta_{m,0}\left (r_{q,\leftarrow}^{(J_1)}\right )^{n+1}+\left (1-\delta_{m,0}\right )\sum_{p=1}^{\min\{m,n+1\}} \binom{m-1}{p-1} \binom{n+1}{p} \left({r_{q,\leftarrow}^{(J_1)}}\right)^{n+1-p}  \left({r_{q,\rightarrow}^{(J_1)}}\right)^{m-p} \left(t_{q,\leftarrow}^{(J_1)} t_{q,\rightarrow}^{(J_1)}\right)^p.
\end{equation}
\end{widetext}
To determine the global reflection coefficient Eq. (\ref{eq:R}), we need to find the explicit expressions of the reflection and transmission coefficients through each barrier. This is done in the next Section.

\subsection{Reflection and transmission coefficients at the barriers}
Here we explicitly determine the coefficients $r_{q,\leftarrow}^{(J_0)}$, $r_{q,\alpha}^{(J_i)}$ and $t_{q,\alpha}^{(J_i)}$ with $\alpha=\leftarrow, \rightarrow$ and $i=1,2$.
Let us consider a junction located at $x=x_0$, which separates regions $A$ (for $x<x_0$, characterized by $K_A$ and $v_A$) and $B$ (for $x>x_0$ characterized by $K_B$ and $v_B$). This situations applies both to $J_1$ ($x_0=l/2$) and $J_2$ ($x_0=L/2$).
The scattering states are
\begin{eqnarray}
f_{q,\rightarrow}&=&\begin{cases}
e^{iq\frac{v_F}{v_A} x} + r_{q,\rightarrow} e^{-iq\frac{v_F}{v_A} x} & x<x_0\nonumber\\
t_{q,\rightarrow} e^{iq\frac{v_F}{v_B} x} & x>x_0
\end{cases}\\\nonumber
f_{q,\leftarrow}&=&\begin{cases}
t_{q,\leftarrow} e^{-iq\frac{v_F}{v_A} x} & x<x_0 \\
e^{-iq\frac{v_F}{v_B} x} + r_{q,\leftarrow} e^{iq\frac{v_F}{v_B} x} & x>x_0\nonumber
\end{cases}
,\end{eqnarray}
corresponding to modes incident from $A$ ($f_{q,\rightarrow}$) and $B$ ($f_{q,\leftarrow}$) respectively.
The continuity of $f_{q,\rightarrow}(x)$ and $\frac{v(x)}{K(x)}f_{q,\rightarrow}(x)$ implies
\begin{subequations}
\label{eq:singleJ->}
\begin{align}
t_{q,\rightarrow} &= \frac{2K_B}{K_A + K_B} e^{iqv_F(\frac{1}{v_A}-\frac{1}{v_B})x_0}\\
r_{q,\rightarrow} &= \frac{K_B-K_A}{K_A+K_B} e^{i2q \frac{v_F}{v_A}x_0}. 
\end{align}
\end{subequations}
In the opposite case,  for scattering from the right the coefficients are
\begin{subequations}
\label{eq:singleJ<-}
\begin{align}
t_{q,\leftarrow} &= \frac{2K_A}{K_A + K_B} e^{iqv_F(\frac{1}{v_A}-\frac{1}{v_B})x_0} \\
r_{q, \leftarrow} &= \frac{K_A-K_B}{K_A+ K_B} e^{-i2q\frac{v_F}{v_B}x_0}.
\end{align}
\end{subequations}
Using the expressions in Eqs. (\ref{eq:singleJ->}a-b) and (\ref{eq:singleJ<-}a-b) it is easy to show that
\begin{align}
\label{eq:coeff_r}
r^{(J_2)}_{q,\leftarrow} &= \frac{K_h - 1}{K_h +1} e^{-iqL} \\
r^{(J_2)}_{q,\rightarrow} &= \frac{1-K_h}{1+K_h} e^{iqL \frac{v_F}{v_h}} \\
t^{(J_2)}_{q,\leftarrow} &= \frac{2 K_h}{K_h +1} e^{iqL(\frac{v_F}{v_h}-1)/2} \\
t^{(J_2)}_{q,\rightarrow} &= \frac{2}{K_h +1} e^{iqL(\frac{v_F}{v_h}-1)/2} \\
r^{(J_1)}_{q,\leftarrow} &= \frac{K_c-K_h}{K_c + K_h} e^{-iql \frac{v_F}{v_h}}\\
r^{(J_1)}_{q,\rightarrow} &= \frac{K_h-K_c}{K_c+K_h} e^{iql \frac{v_F}{v_c}} \\
t^{(J_1)}_{q,\leftarrow} &= \frac{2 K_c}{K_c +K_h} e^{iql(\frac{v_F}{v_c}-\frac{v_F}{v_h})/2} \\
t^{(J_1)}_{q,\rightarrow} &= \frac{2 K_h}{K_c +K_h} e^{iql(\frac{v_F}{v_c}-\frac{v_F}{v_h})/2}.
\end{align}
In addition, focusing on the scattering state incoming on $J_0$ and requiring that the current vanishes at $x=0$ one finds
\begin{equation}\label{eq:coeff_r_end}
r^{(J_0)}_{q,\leftarrow}=-1
.\end{equation}
Substituting Eqs. (\ref{eq:coeff_r}-\ref{eq:coeff_r_end}) into Eqs. (\ref{eq:R}) and (\ref{eq:Pnm})) we obtain
\begin{widetext}
\begin{equation}
R_q	 = \frac{K_h - 1}{K_h +1} e^{-iqL} + \frac{4 K_h}{(K_h +1)^2} e^{iqL(\frac{v_F}{v_h}-1)} \sum_{n,m=0}^{\infty} (-1)^m \left(\frac{1-K_h}{1+K_h}\right)^n  e^{iqL \frac{v_F}{v_h} n} \; P_{n,m} 
\end{equation}
and
\begin{equation}
\begin{split}
P_{n,m} & = \delta_{m,0}\left(\frac{K_c-K_h}{K_c + K_h}\right)^{n+1} e^{-iql \frac{v_F}{v_h} (n+1)} \\ & +(1-\delta_{m,0})e^{-ikl \frac{v_F}{v_h} (n+1)} e^{ikl \frac{v_F}{v_c}(m)} \left(\frac{K_c-K_h}{K_c + K_h}\right)^{m+n+1} (-1)^m \sum_{p=1}^{\min\{m,n+1\}} \binom{m-1}{p-1} \binom{n+1}{p} \left(\frac{ - 4 K_c K_h}{(K_c - K_h)^2}\right)^p.
\end{split}
\end{equation}
Using straightforward algebra the global reflection coefficient can be expressed in compact form as
\begin{equation}
\label{eq:R_def}
R_q = \frac{K_h - 1}{K_h +1} e^{-iqL} + \frac{4 K_h}{(K_h +1)^2} \sum_{n,m} C_{n,m} \,\exp\left[iq \left((L-l)\frac{v_F}{v_h}(n+1) + l \frac{v_F}{v_c} m - L \right)\right]
\end{equation}
with
\begin{equation}\label{eq:suppAc}
C_{n,m}=\left (\frac{1-K_h}{1+K_h}\right )^n\left (\frac{K_c-K_h}{K_c+K_h}\right )^{n+m+1}\left [(1-\delta_{m,0})\sum_{p=1}^{\mathrm{min}\{m,n+1\}}{{m-1}\choose{p-1}}{{n+1}\choose{p}}\left [\frac{-4 K_hK_c}{\left (K_c-K_h\right )^2}\right ]^p+\delta_{m,0}\right ]
.\end{equation}
Finally, by restoring the index $\lambda=c$ for the charge sector, using Eq. (\ref{eq:riflesso}) we find
\begin{equation}\label{eq:supprho(x,t)}
\rho^{(refl)}_c(x_d,t)=r\rho^{(0)}(-x_d+v_Ft+L)+\gamma\sum_{n,m=0}^{\infty}C_{n,m}\rho^{(0)}\left (-x_d+v_Ft+L-(n+1)\frac{v_F}{v_h}(L-l)-m\frac{v_F}{v_c}l\right )
,\end{equation}
with $r=\frac{1-K_h}{1+K_h}$, $\gamma=-\frac{4K_h}{(1+K_h)^2}$. The spin density evolution $\rho_s(x_d,t)$ can be obtained from Eqs. (\ref{eq:suppAc}-\ref{eq:supprho(x,t)}) with the substitutions $\left (v_c,K_c,K_h\right )\to\left (v_s,K_s,1/K_h\right )$.
\end{widetext}


\begin{thebibliography}{43}%
\bibitem{giamarchi2003quantum} T. Giamarchi, \textit{Quantum Physics in One Dimension} (Oxford University Press, Oxford, 2003).
\bibitem{deshpande2010electron}V. V. Deshpande, M. Bockrath, L. I. Glazman, and A. Yacoby, Nature {\bf 464}, 209 (2010).
\bibitem{barak2010interacting} G. Barak, H. Steinberg, L. N. Pfeiffer, K. W. West, L. Glazman, F. Von Oppen, and A. Yacoby, Nat. Phys. {\bf 6}, 489 (2010).
\bibitem{pham2000fractional}K.-V. Pham, M. Gabay, and P. Lederer, Phys. Rev. B {\bf 61}, 16397 (2000).
\bibitem{maslov1995landauer}D. L. Maslov and M. Stone, Phys. Rev. B {\bf 52}, 5539(R) (1995).
\bibitem{safi1995transport}I. Safi and H. Schulz, Phys. Rev. B {\bf 52}, 17040(R) (1995).
\bibitem{auslaender01042005} O. M. Auslaender, H. Steinberg, A. Yacoby, Y. Tserkovnyak, B. I. Halperin, K. W. Baldwin, L. N. Pfeiffer, and K. W. West, Science {\bf 308}, 88 (2005).
\bibitem{Lorenz2002evidence}
T. Lorenz, M. Hofmann, M. Gr\"uninger, A. Freimuth, G. S. Uhrig, M. Dumm, and M. Dressel, 
Nature {\bf 418}, 614 (2002).
\bibitem{bockrath1999luttinger}
M. Bockrath, D. H. Cobden, J. Lu, A. G. Rinzler, R. E. Smalley, L. Balents, and P. L. McEuen, Nature {\bf 397}, 598 (1999).
\bibitem{steinberg2007charge} H. Steinberg, G. Barak, A. Yacoby, L. N. Pfeiffer, K. W. West, B. I. Halperin, and K. Le Hur, Nature Physics {\bf 4}, 116 (2007).
\bibitem{kamata2014fractionalized}H. Kamata, N. Kumada, M. Hashisaka, K. Muraki, and T. Fujisawa, Nature NanoTech. {\bf 9}, 177 (2014).
\bibitem{tomonaga1950remarks} S.-I. Tomonaga, Prog. Thero. Phys. {\bf 5}, 544 (1950).
\bibitem{luttinger1963exactly} J. M. Luttinger, J. Math. Phys. {\bf 4}, 1154 (1963).
\bibitem{tarucha1995reduction}S. Tarucha, T. Honda, and T. Saku, Sol. Stat. Comm. {\bf 94}, 413 (1995).
\bibitem{safi1997properties} I. Safi, Ann. Phys. (Paris) {\bf 22}, 463 (1997).
\bibitem{bena2001measuring} C. Bena, S. Vishveshwara, L. Balents, and M. P. A. Fisher, J. Stat. Phys. {\bf 103}, 429 (2001).
\bibitem{trauzettel2004appearance} B. Trauzettel, I. Safi, F. Dolcini, and H. Grabert, Phys. Rev. Lett. {\bf 92}, 226405 (2004).
\bibitem{berg2009fractional} E. Berg, Y. Oreg, E.-A. Kim, and F. Von Oppen, Phys. Rev. Lett. {\bf 102}, 236402 (2009).
\bibitem{le2008charge}K. Le Hur, B. I. Halperin, and A. Yacoby, Ann. Phys. {\bf 323}, 3037 (2008).
\bibitem{perfetto2014time}E. Perfetto, G. Stefanucci, H. Kamata, and T. Fujisawa, Phys. Rev B {\bf 89}, 201413(R) (2014).
\bibitem{roth2009nonlocal}A. Roth, C. Brune, H. Buhmann, L. W. Molenkamp, J. Maciejko, X.-L. Qi, and S.-C. Zhang, Science {\bf 325}, 294 (2009).
\bibitem{hasan2010colloquium}M. Z. Hasan and C. L. Kane, Rev. Mod. Phsy. {\bf 82}, 3045 (2010).
\bibitem{qi2011topological}X.-L. Qi and S.-C. Zhang, Rev. Mod. Phys. {\bf 83}, 1057 (2011).
\bibitem{wu06}C. Wu, B. A. Bernevig, and S.-C. Zhang, Phys. Rev. Lett. {\bf 96}, 106401 (2006).
\bibitem{dolcini11}
F. Dolcini, Phys. Rev. B {\bf 83}, 165304 (2011).
\bibitem{dolcini12} F. Dolcini, Phys. Rev. B {\bf 85}, 033306 (2012).
\bibitem{romeo14}
F. Romeo and R. Citro, Phys. Rev. B {\bf 90}, 155408 (2014). 
\bibitem{dolcetto2012tunneling} G. Dolcetto, S. Barbarino, D. Ferraro, N. Magnoli, and M. Sassetti, Phys. Rev. B {\bf 85}, 195138 (2012).
\bibitem{dolcetto13} G. Dolcetto, F. Cavaliere, D. Ferraro, and M. Sassetti, Phys. Rev. B {\bf 87}, 085425 (2013).
\bibitem{das2011spin}S. Das and S. Rao, Phys. Rev. Lett. {\bf 106}, 236403 (2011).
\bibitem{Hou2009cornet} C.-Y.Hou, E.-A.Kim, and C. Chamon, Phys. Rev. Lett. {\bf 102}, 076602 (2009).
\bibitem{Schmidt2011current} T. L. Schmidt, Phys. Rev. Lett. {\bf 107}, 096602 (2011).
\bibitem{Budich2012phonon} J. C. Budich, F. Dolcini, P. Recher, and B. Trauzettel, Phys. Rev. Lett. {\bf 108}, 086602 (2012).
\bibitem{garate2012noninvasive}I. Garate and K. Le Hur, Phys. Rev. B {\bf 85}, 195465 (2012).
\bibitem{Li2015} T. Li, P. Wang, H. Fu, L. Du, K. A. Schreiber, X. Mu, X. Liu, G. Sullivan, G. A. Csáthy, X. Lin, R.-R. Du. arXiv/Cond. Mat:1507.08362 (2015).
\bibitem{calzona2015transient} A. Calzona, M. Carrega, G. Dolcetto, and M. Sassetti, Physica E {\bf 74}, 630 (2015).
\bibitem{ashoori1992edge}R. Ashoori, H. Stormer, L. N. Pfeiffer, K. Baldwin, and K. W. West, Phys. Rev. B {\bf 45}, 3894 (1992).
\bibitem{kumada2014intrinsic}N. Kumada, P. Roulleau, B. Roche, M. Hashisaka, H. Hibino, I. Petkovic, and D. C. Glattli, ArXiv/Cond. Mat:1407.4379 (2014).
\bibitem{waldie2015measurement}J. Waldie, P. See, V. Kashcheyevs, J. Griffiths, I. Farrer, G. Jons, D. Ritchie, T. Janssen, and M. Kataoka, ArXiv/Cond. Mat:1503.07140 (2015).
\bibitem{g-ology} In this sense, the meaning of the parameters $g_{i,\parallel(\perp)}$ becomes clear in the context of $g$-ology~\cite{giamarchi2003quantum}.
\bibitem{Agarwal2014time} A. Agarwal, Phys. Rev. B {\bf 90}, 195403 (2014).
\bibitem{ferraroprl14}
D. Ferraro, B. Roussel, C. Cabart, E. Thibierge, G. F\`eve, C. Grenier, and P. Degiovanni, Phys. Rev. Lett. {\bf113}, 166403 (2014).
\bibitem{bocquillon2012electron}E. Bocquillon, F. Parmentier, C. Grenier, J.-M. Berroir, P. De Giovanni, D.C. Glattli, B. Placais, A. Cavanna, Y. Jin, and G. Feve, Phys. Rev. Lett. {\bf 108}, 196803 (2012).

\bibitem{bocquillon2014electron}E. Bocquillon, V. Freulon, F.D. Parmentier, J.-M- Berroir, B. Pla\c cais, C. Wahl, J. Rech, T. Jonckheere, T. Martin, C. Grenier, D. Ferraro, P. Degiovanni, and G. F\`eve, Ann. Phys. (Berlin) {\bf 526}, 1 (2013). 
\bibitem{wahl13}
C. Wahl, J. Rech, T. Jonckheere, and T. Martin, Phys. Rev. Lett. 112, 046802 (2014).
\bibitem{ferraro14}
D. Ferraro, C. Wahl, J. Rech, T. Jonckheere, and T. Martin, Phys. Rev. B {\bf 89}, 075407 (2014). 
\bibitem{dolcini2005transport}F. Dolcini, B. Trauzettel, I. Safi, and H. Grabert, Phys. Rev. B {\bf 71}, 165309 (2005).
\bibitem{note} In the physical 2-channel scenario of Fig. \ref{fig:1}(b) this condition implies that no barrier exists at $x=0$ and incoming wave-packets are fully transmitted.
\bibitem{note2} The presence of the gate electrode is expected to screen the strength of electron interactions, leading to $K_h < K_c <1$ and $1< K_s < 1/K_h$.
\bibitem{vayrynen2013helical}J. I. Vaeyrynen, M. Goldstein, and L. I. Glazman, Phys. Rev. Lett. {\bf 110}, 216402 (2013).
\bibitem{vayrynen2014resistance}J. I. Vaeyrynen, M. Goldstein, Y. Gefen, and L. I. Glazman, Phys. Rev. B {\bf 90}, 115309 (2014).
\bibitem{konig2007quantum}M. Koenig, S. Videmann, C. Brune, A. Roth, H. Buhmann, L. W. Molenkamp, X.-L. Qi, and S.-C. Zhang, Science {\bf 318}, 766 (2007).
\bibitem{konig2008quantum}M. Koenig, H. Buhmann, L. W. Molenkamp, T. Hughes, C.-X. Liu, x.-L. Qi, and S.-C. Zang, J. Phys. Soc. Jap {\bf 77}, 031007 (2008).
\bibitem{konig2013spatially}M. Koenig, M. Baenninger, A. G. Garcia, N. Harjee, B. L. Pruitt, C. Ames, P. Leubner, C. Bruune, H. Buhmann, L. W. Molenkamp, and D. Goldhaber-Gordon, Phys. Rev. X {\bf 3}, 021003 (2013).
\bibitem{gusev2014temperature} G. Gusev, Z. Kvon, E. Olshanetsky, A. Levin, Y. Krupko, J. Portal, N. Mikhailov, and S. Dvoretski, Phys. Rev. B {\bf 89}, 125305 (2014).
\bibitem{olshanetsky2015persistence} E. Olshanetsky, Z. Kvon, G. Gusev, A. Levin, O. Raichev, N. Mikhailov, and S. Dvoretsky, Phys. Rev. Lett. {\bf 114}, 126802 (2015).
\bibitem{essert2015two} S. Essert, V. Krueckl, and K. Richter, ArXiv:CondMat/1507.00928 (2015).
\end{thebibliography}
\end{document}